%% file: acl_latex.tex
\pdfoutput=1

\documentclass[11pt]{article}

\usepackage[final]{acl}

\usepackage{times}
\usepackage{latexsym}

\usepackage[T1]{fontenc}

\usepackage[utf8]{inputenc}

\usepackage{microtype}

\usepackage{inconsolata}

\usepackage{graphicx}
\usepackage{amsmath}
\usepackage{todonotes}
\usepackage{xcolor}         
\usepackage{amssymb}
\usepackage{algorithm}
\usepackage{algorithmic}
 
\title{Evaluation on Entity Matching in Recommender Systems}

\author{Zihan Huang, Rohan Surana, Zhouhang Xie, Junda Wu, Yu Xia, Julian McAuley \\
        University of California, San Diego}

\usepackage{booktabs}

\usepackage{graphicx}
\usepackage{subcaption}

\usepackage{tabularx}

\begin{document}
\maketitle

\input{contents/0_abstract}

\input{contents/1_intro}

\input{contents/2_related}

\input{contents/3_prelim}
\input{contents/4_exp}
\input{contents/6_conclusion}

\bibliography{custom}

\appendix

\end{document}

%% file: contents/0_abstract.tex
\begin{abstract}

Entity matching is a crucial component in various recommender systems, including conversational recommender systems (CRS) and knowledge-based recommender systems. However, the lack of rigorous evaluation frameworks for cross-dataset entity matching impedes progress in areas such as LLM-driven conversational recommendations and knowledge-grounded dataset construction. 

In this paper, we introduce Reddit-Amazon-EM, a novel dataset comprising naturally occurring items from Reddit and the Amazon '23 dataset. Through careful manual annotation, we identify corresponding movies across Reddit-Movies and Amazon'23, two existing recommender system datasets with inherently overlapping catalogs. Leveraging Reddit-Amazon-EM, we conduct a comprehensive evaluation of state-of-the-art entity matching methods, including rule-based, graph-based, lexical-based, embedding-based, and LLM-based approaches.

For reproducible research, we release our manually annotated entity matching gold set and provide the mapping between the two datasets using the best-performing method from our experiments. This serves as a valuable resource for advancing future work on entity matching in recommender systems. Data and Code are accessible at: \href{https://github.com/huang-zihan/Reddit-Amazon-Entity-Matching}{https://github.com/huang-zihan/Reddit-Amazon-Entity-Matching}.
\end{abstract}

%% file: contents/1_intro.tex
\section{Introduction}

Entity Matching (EM) has become increasingly essential and is utilized in various aspects of recommender systems, particularly in tasks involving integration across diverse data formats, such as Conversational Recommendation Systems (CRS) and recommendation dataset linking. CRS requires mapping items mentioned in content generated by LLMs into items in a structured catalogue. Building datasets in recommendation fields frequently requires linking entities from the wild database with different signals, such as linking data that comprises movie descriptions with movie ratings, with different representations of the same movie items.

In LLM-based CRS, for example, we employ LLMs to recommend movies for users through a conversation. For reasons such as ensuring the reliability of LLM recommendations, EM methods that swiftly and accurately match items mentioned in LLM responses into a movie database are needed. Recent works in CRS~\cite{DBLP:conf/cikm/HeXJSLFMKM23, gao2021advances, 10.1145/3583780.3614949} employ EM methods such as Fuzzy, BM25, Faiss, etc. However, none of these studies rigorously evaluate the performance of EM, which is crucial for the recommendation system when employing them. Additionally, there is no consensus in the literature regarding which EM methods are most effective for tasks such as CRS and data linking, leaving this as an open problem.

In this work, we study the methods that rigorously evaluate EM methods through constructing an EM dataset, specifically in the scenarios of CRS, due to EM methods' frequent and critical usage in CRS. The traditional construction of EM datasets in CRS involves collecting items with structured knowledge, and hiring crowd-workers to generate conversations, role-playing item seeker and recommender~\cite{li2018conversational, liu-etal-2020-towards-conversational, liu-etal-2021-durecdial, DBLP:conf/recsys/ManzoorJ22a}. However, recent studies highlight the advantages of leveraging large volumes of naturally occurring user conversations, which often feature semantically rich queries~\cite{DBLP:conf/cikm/HeXJSLFMKM23}. Unfortunately, these in-the-wild queries typically lack associated structured information about items, hindering the development of scalable, knowledge-aware CRS models that can effectively utilize such user interactions~\citep{hayati2020inspired,manzoor2022inspired2,li2018towards,liu2020towards}. While there have been attempts to link items from in-the-wild conversations, such as those from Reddit, to knowledge bases like IMDb~\cite{maas-EtAl:2011:ACL-HLT2011}, the methodologies for ensuring accurate item matching across datasets remain unclear~\cite{rabiah2024bridging}. 

To address these challenges, we present Reddit-Amazon-EM, the largest publicly available knowledge-grounded EM dataset to date, where movies discussed in Reddit conversations are syntactically matched to corresponding Amazon entities. This dataset includes various naturally occurring movies and structured
item knowledge. We rigorously benchmark existing EM methods for cross-dataset item matching, particularly focusing on movies, and carefully select effective techniques for linking entities across diverse datasets. Our work not only facilitates reproducible research but also establishes a robust evaluation framework for assessing entity matching performance, thereby bridging the gap between real-world user interactions and structured knowledge integration.

Our contributions are as follows:
\begin{itemize}
\item We present Reddit-Amazon-EM, a carefully annotated dataset of over 4k items mapping between two in-the-wild platforms sharing overlapping entities.
\item We rigorously evaluate entity matching baselines with our human-annotated dataset, demonstrating that hybrid methods combining structural and semantic signals consistently outperform standalone classic retrieval methods both on the Reddit-Amazon-EM dataset and downstream tasks.
\item We make our human-annotated dataset and evaluation code available to facilitate EM evaluation for future research, such as LLM-driven conversational recommendations and knowledge-grounded dataset construction.
\end{itemize}

%% file: contents/2_related.tex
\section{Related Work}

Entity matching identifies records referencing identical real-world entities across datasets~\cite{10.1145/3442200}. Traditional EM methods depend on handcrafted features, such as Jaro-Winkler similarity, or probabilistic models, but face scalability challenges with noisy, heterogeneous data. Recent breakthroughs leverage transformer architectures such as BERT~\cite{devlin-etal-2019-bert}, GNEM~\cite{10.1145/3442381.3450119}, and graph neural networks~\cite{10386187} to capture semantic relationships beyond syntactic patterns, while cross-attention mechanisms model interdependencies between records~\cite{10.1007/s00778-023-00779-z, 10.1145/3431816}. However, consistent evaluation of EM performance across domains remains unresolved, particularly for large-scale real-world applications.

EM are frequently employed in CRS. Current CRS have transformed static recommendation paradigms into dynamic, multi-turn dialogues~\cite{zaidi2024review}, integrate NLP to interpret open-ended queries~\cite{10.1145/3240323.3240394}, contextual prompting for intent alignment and unify knowledge-enhanced prompts~\cite{10.1145/3534678.3539382}. The inherent randomness and vagueness in conversation makes grounding conversational queries into structured catalogue with effective EM a persistent challenge. By introducing the Reddit-Amazon-EM benchmark with annotated mappings between Reddit and Amazon movies, we enable systematic assessment of EM methods, bridging the divide between EM techniques and LLM-augmented CRS, fostering scalable, knowledge-aware recommender systems with reproducible evaluation protocols.

%% file: contents/3_prelim.tex
\section{Entity Matching Task Formulation}

EM associates ambiguous textual mentions of entities such as product names, movies, or personas in conversational queries, with corresponding canonical representations in a structured knowledge base.

In our context, EM maps Reddit conversations to Amazon Movie knowledge base entries. Let $\mathcal{M} = \{m_1, \ldots, m_N\}$ denote the set of entity mentions extracted from Reddit conversations, where each $m_i \in \mathcal{M}$ corresponds to an ambiguous textual reference to movie entities. Given a structured knowledge base $\mathcal{K} = \{e_1, \ldots, e_M\}$ containing canonical Amazon Movie entries 
we formulate the entity linking task as:

\[
f: \mathcal{M} \times \mathcal{C}^k \rightarrow \mathcal{K} \cup \{\text{NIL}\}
\]

where $\mathcal{C}^k = \{c_1, \ldots, c_k\} \subseteq \mathcal{K}$ denotes the candidate set retrieved through blocking methods such as fuzzy matching for mention $m_i$. NIL denotes unlinkable items.

In Reddit-Amazon-EM evaluation, we formalize entity matching as a binary classification task over candidate mention-entry pairs. Let $\mathcal{D} = \{(m_i, e_j, y_{ij})\}$ denote our annotated dataset where $y_{ij} \in \{0,1\}$ indicates whether mention $m_i \in \mathcal{M}$ corresponds to knowledge base entry $e_j \in \mathcal{K}$. For each candidate pair $(m_i, e_j) \in \mathcal{M} \times \mathcal{C}^k$, we define a scoring function $s: \mathcal{M} \times \mathcal{K} \rightarrow [0,1]$ that computes similarity between textual mentions and structured entries. The binary classification decision is then made through thresholding:

\[
\hat{y}_{ij} = 
\begin{cases} 
1 & \text{if } \sigma(s(m_i, e_j)) \geq \tau \\
0 & \text{otherwise}
\end{cases}
\]

where $\sigma(\cdot)$ denotes score normalization across candidate pairs for mention $m_i$, and $\tau \in [0,1]$ is a tunable decision threshold.

\section{Data Collection and Human Annotation for Reddit-Amazon EM}

\begin{figure*}[h]
    \centering
    \includegraphics[width=1\linewidth]{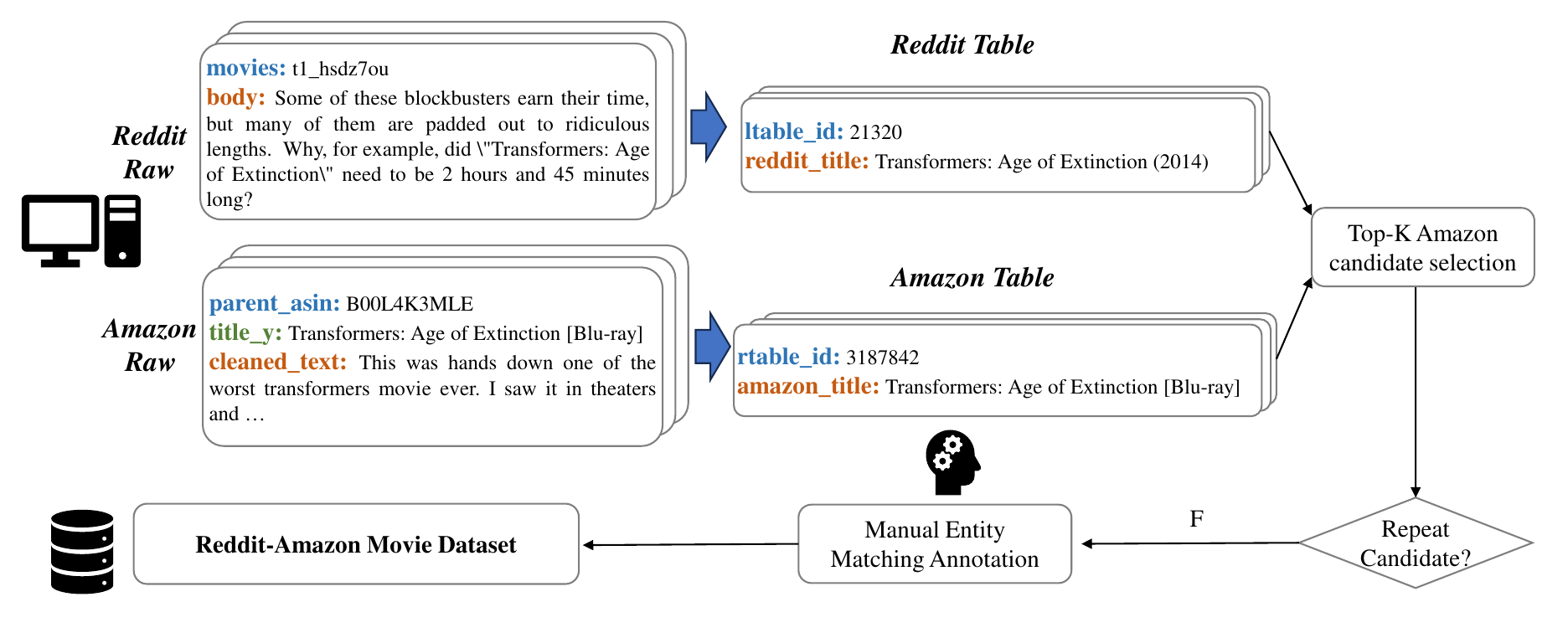}
    \caption{Reddit-Amazon-EM Dataset Construction.}
    \label{fig:enter-label}
    
\end{figure*}

\begin{table*}[h]
\centering
\caption{Dataset Samples of Reddit-Amazon-EM Dataset}
\label{tab:benchmark_samples}
\begin{tabularx}{\textwidth}{@{} lllllc @{}}
\toprule
\multicolumn{1}{c}{\textbf{ltable\_id}} & 
\multicolumn{1}{c}{\textbf{rtable\_id}} & 
\multicolumn{1}{c}{\textbf{reddit\_title}} & 
\multicolumn{1}{c}{\textbf{amazon\_title}} & 
\multicolumn{1}{c}{\textbf{parent\_asin}} & 
\multicolumn{1}{c}{\textbf{label}} \\
\midrule
14938 & 3830856 & Seasons (2015) & Seasons (Large Format) & B0000694Z1 & 1 \\
21885 & 5914787 & Videodrome (1983) & Videos & B00008W2RE & 0 \\
19548 & 265388 & The Package (2018) & Terror on Tape & B000MOKVRG & 0 \\
9404 & 4772763 & Kedi (2016) & Jet Li & B0091JJ2NW & 0 \\
21060 & 4638912 & Time Walker (1982) & The Horses of McBride & B00EJE4DG0 & 0 \\
15549 & 2984641 & Sliders (2018) & Crunch: Yoga Body Sculpt & B001C0NMUW & 0 \\
977 & 2838303 & Alien (1979) & Alien [Blu-ray] & B00MBNZER8 & 1 \\
11368 & 4589640 & Mirrormask (2005) & All American Bully & B00U2YNN0I & 0 \\
\bottomrule
\end{tabularx}
\end{table*}

\begin{table}[h]
\centering
\caption{Dataset Statistics}
\label{tab:dataset_stats}
\begin{tabular}{l@{\hspace{50pt}}r}
\toprule
\textbf{Statistic} & \textbf{Count} \\
\midrule
Total           & 47,070 \\ 
Train           & 30,124 \\
Validation      & 7,532  \\
Test            & 9,414  \\
Positive        & 4,322  \\
Negative        & 42,748 \\
\bottomrule
\end{tabular}
\end{table}

We construct the Reddit-Amazon-EM dataset following the pipeline in Figure~\ref{fig:enter-label}. We extract approximately 1,000 frequent movie titles from a public Reddit conversational dataset~\cite{10.1145/3583780.3614949}. To address metadata discrepancies between Reddit's informal content and Amazon's structured catalog~\cite{hou2024bridging}, we implement a two-stage process combining automated candidate retrieval followed by a rigorous human annotation process to finalize accurate entity correspondences, as shown in the bottom right corner of Figure~\ref{fig:enter-label}.

First, for each Reddit movie title, we retrieve the ten most relevant candidates from the Amazon dataset using title-based similarity metrics (minimum edit distance and embedding similarity). We apply metadata filtering to exclude candidates with mismatched release years or television indicators such as season numbers.

Subsequently, we perform detailed human annotation using an interactive Streamlit-based interface. Annotators review each Reddit movie alongside supporting metadata, including minimum edit distance, the nearest Amazon title according to this metric, a filtered candidate list, and additional suggestions generated by prompting GPT-3.5. Annotators then manually select all semantically correct matches. For example, the Reddit title \textit{“Prisoners (2013)”} was evaluated against Amazon candidates like \textit{“Prisoners (Blu-ray+DVD)”} and \textit{“Prisoners [DVD] (2013)”}, while clearly unrelated items such as \textit{“PRISONER”} or \textit{“Prison (Collector’s Edition)”} were rejected. After completing the annotation, we retained 869 Reddit titles confirmed to have accurate Amazon correspondences.

Table~\ref{tab:human_annotation_stats_horizontal} summarizes the outcomes of this annotation process. Of the 868 annotated Reddit movie titles, 820 titles successfully matched at least one Amazon movie, while 48 titles had no suitable matches. On average, each Reddit movie aligned with 5.50 Amazon candidates when considering all annotated titles, and 5.82 candidates for movies with at least one confirmed match. In total, annotators identified 4,504 unique Amazon movie entities as matches. These statistics highlight both the diversity inherent in user-generated Reddit content and the complexity involved in accurately linking such social media mentions to structured catalog entries.

For evaluating baseline entity matching methods described in the subsequent section, namely GNEM and ComEM, we utilize the human-annotated dataset to construct positive and negative training pairs. Positive pairs are directly derived from confirmed human matches, yielding a total of 4,322 pairs. To ensure a robust evaluation, we significantly expand our negative training set by incorporating two complementary sources: candidate-based negatives, comprising Amazon items explicitly rejected by annotators, and additional randomly-generated negative pairs linking Reddit titles with unrelated Amazon entities. Specifically, we set the random negative multiplier to 9, maintaining approximately a 1:10 positive-to-negative ratio. This procedure results in a comprehensive dataset consisting of 4,322 positive and 42,748 negative pairs, detailed further in Table~\ref{tab:dataset_stats}.

This carefully designed annotation and dataset construction approach provides a reliable and comprehensive foundation for rigorous evaluation and effective modeling of entity matching methods within recommender system research.

\begin{table*}[ht]
\centering
\caption{Statistics on Reddit-Amazon-EM Human-Annotation}
\label{tab:human_annotation_stats_horizontal}
\resizebox{\textwidth}{!}{%
\begin{tabular}{lcccccc}
\toprule
\textbf{Statistic} & 
\textbf{Total Reddit titles} & 
\textbf{Titles with $\geq$1 match} & 
\textbf{Titles with no match} & 
\textbf{Avg. matches (all)} & 
\textbf{Avg. matches (matched)} & 
\textbf{Unique match candidates} \\
\midrule
\textbf{Value} & 
868 & 
820 & 
48 & 
5.502 & 
5.824 & 
4,504 \\
\bottomrule
\end{tabular}
}
\end{table*}

%% file: contents/4_exp.tex
\section{Experiments and Results}

\subsection{Experimental Setting}

We split the dataset containing manually annotated positive links and negatively sampled pairs into training (30,124), validation (7,532), and test (9,414) samples to ensure robust model training and comprehensive evaluation.

We present the dataset statistics used in our experiments in Table~\ref{tab:dataset_stats}. Each sample contains 6 attributes: 

\begin{itemize}
    \item \textbf{ltable\_id}: The index of query item in Reddit.
    \item \textbf{rtable\_id}: The index of target movie title in Amazon datasets.
    \item \textbf{reddit\_title}: Texts content of Reddit query.
    \item \textbf{amazon\_title}: Title of the item in Amazon.
    \item \textbf{parent\_asin}: Parent ID in Amazon.
    \item \textbf{label}: Denote whether these two items are a correct match.
\end{itemize}

We evaluate the following EM methods on Reddit-Amazon-EM.

\noindent\textbf{BM25} We adopt Okapi BM25 as a lexical-based EM baseline, which utilizes term frequency and inverse document frequency to estimate relevance.

\noindent\textbf{Faiss} We utilize Faiss~\cite{douze2025faisslibrary} as a dense vector retrieval baseline, with query and items encoded with all-MiniLM-L6-v2~\citep{wang2020minilm} and nearest-neighbor search performed using inner-product similarity (IndexFlatIP).

\noindent\textbf{Embedding + Fuzzy} We evaluated a hybrid matching method that use Bert embeddings concatenated with three fuzzy matching metrics (Levenshtein edit ratio, Jaro-Winkler similarity, and Jaccard token overlap).

\noindent\textbf{GNEM} We evaluate GNEM (Graph Neural Entity Matching)~\citep{10.1145/3442381.3450119} as a graph-based baseline, with a weighted record-pair graph where edges represent cross-instance relationships and a single-layer gated graph convolution network for graph propagation.

\noindent\textbf{ComEM} We adapt ComEM~\citep{wang-etal-2025-match} to one-to-many through as an LLM-based EM baseline with a two-phase candidate retrieval and selection pipeline.

We use the following evaluation metrics for EM baselines on Reddit-Amazon-EM and downstream tasks. Specifically, Recall@k measures the proportion of relevant entities retrieved among the top-\textit{k} candidates, and additionally, Precision@k, F1, and accuracy score.

\subsection{Evaluation Results and Main Findings}

Our experiments reveal a clear hierarchy in model effectiveness in Table~\ref{tab:reddit_amazon_bert}. Graph-based GNEM achieves state-of-the-art performance with a F1 of 96.29\% and an Accuracy of 96.74\%, significantly outperforming traditional baselines BM25 and Faiss by 17.9–24.0\% absolute F1 gain. The performance gap benefits from GNEM’s capability of distinguishing similar movie titles with different release years and product formats such as "Prisoners (Blu-ray+DVD)" and "Prisoners [DVD] (2013)", and its effectiveness in semantic matching based on graph-based architecture.

ComEM, leveraging LLM-enhanced retrieval, follows closely with a F1 of 94.02\%, also significantly outperforms traditional baselines. The LLM-powered semantic understanding and knowledge enhancement enables ComEM to distinguish literally similar but semantically different movie titles. However, its lower precision relative to GNEM suggests LLMs’ weaker grasp of details in movie titles compared to symbolic rules, such as exact numeric identifiers.

The Embedding+Fuzzy hybrid approach achieves an F1 of 86.68\%, demonstrating the complementary strengths of neural embeddings and symbolic matching. In contrast, standalone traditional methods such as BM25 and Faiss reveal characteristic trade-offs. When choosing the threshold with the highest F-1 score, Faiss achieves a high recall of 89.76\% but suffers from a low precision 60.51\%, frequently retrieving semantically related but incorrect candidates. While BM25 exhibits more balanced and a higher F1 of 78.43\%, but struggles with literal variations.

Table~\ref{tab:time_comparison} compares the computational efficiency across different methods on A6000 GPUs and EPYC 7702 CPUs. Traditional EM methods such as BM25 and Faiss that run on CPUs demonstrate efficient initialization but suffer from prohibitively slow inference due to CPU-based operations. In contrast, modern neural-based EM methods including GNEM and ComEM require substantial training time for graph optimization or LLM pretraining but delivers efficient inference. The hybrid Emb+Fuzzy approach strikes a notable balance, offering fast training and inference.

\begin{table}
\centering
\caption{Computational Time Comparison (30k training/initialization samples, 9.4k test samples)}
\label{tab:time_comparison}
\begin{tabular}{l@{\hspace{8pt}}c@{\hspace{10pt}}c}
\toprule
\textbf{Method} & \textbf{Training/Initialization} & \textbf{Inference} \\
\midrule
GNEM            & 423s/epoch*10    & $\sim$60s          \\
ComEM           & N/A                             & $\sim$70s          \\
Emb+Fuzzy & 30s/epoch*10  & $\sim$10s          \\
BM25            & 16s                             & 8-10h              \\
Faiss           & 104s                            & 8-10h              \\
\bottomrule
\end{tabular}
\end{table}

\begin{table*}[h]
\centering
\caption{EM methods' performance on Reddit-Amazon-EM with Bootstrap SEM 1000 resample.}
\begin{tabular}{lcccc}
\toprule
\textbf{Model} & \textbf{Precision} & \textbf{Recall@1} & \textbf{F1 score} & \textbf{Accuracy} \\
\midrule
Emb+Fuzzy & 86.38 ± 0.03 & 86.99 ± 0.04 & 86.68 ± 0.03 & 92.78 ± 0.02 \\
BM25 & 74.93 ± 0.04 & 82.30 ± 0.04 & 78.43 ± 0.03 & 89.71 ± 0.02 \\
Faiss & 60.51 ± 0.04 & 89.76 ± 0.03 & 72.28 ± 0.04 & 91.83 ± 0.02 \\
BM25 + Faiss & 74.54 ± 0.04 & 84.49 ± 0.04 & 79.20 ± 0.03 & 90.75 ± 0.02 \\
ComEM & 94.50 ± 0.04 & 93.97 ± 0.04 & 94.02 ± 0.04 & 94.70 ± 0.04 \\
GNEM & \textbf{95.82 ± 0.02} & \textbf{96.78 ± 0.02} & \textbf{96.29 ± 0.01} & \textbf{96.74 ± 0.01} \\
\bottomrule
\end{tabular}
\label{tab:reddit_amazon_bert}
\end{table*}

\section{The Impact of EM in Recommender System: a Case Study on CRS}

In this section, we demonstrate the impact of entity matching in recommender systems through a case study of different EM methods for LLM-based conversational recommendation~\cite{DBLP:conf/cikm/HeXJSLFMKM23}. Unlike traditional CRS, which provides static suggestions based on structured user data, LLM-based CRS delivers recommendations through interactive, natural language dialogues with users. This approach enables dynamic personalization, real-time adaptation to user feedback, and improved explainability through conversational context, presenting several advantages over traditional recommendation methods. We evaluate each entity matching method's ability to retrieve relevant movies from structured databases within human-LLM conversational contexts. Table~\ref{tab:retrieval} presents the Recall@1 and Recall@5 metrics across four LLM recommendation responses, namely GPT4~\cite{DBLP:journals/corr/abs-2303-08774}, GPT3.5-turbo~\citep{openai2023gpt35turbo}, Qwen3-4b~\cite{qwen3} and Phi3-mini~\citep{abdin2024phi}.

\begin{table*}[ht]
\centering
\caption{EM methods' performance on CRS dialogue with Bootstrap SEM 1000 resample.}
\label{tab:retrieval}
\begin{tabular}{@{}l *{4}{@{\hspace{3pt}}cc} @{}}
\toprule
 & \multicolumn{2}{c}{GPT4} & \multicolumn{2}{c}{GPT3.5} & \multicolumn{2}{c}{Qwen3} & \multicolumn{2}{c}{Phi3} \\
\cmidrule(lr){2-3} \cmidrule(lr){4-5} \cmidrule(lr){6-7} \cmidrule(lr){8-9}
Method & R@1 & R@5 & R@1 & R@5 & R@1 & R@5 & R@1 & R@5 \\
\midrule
BM25    & {2.08±.003} & 6.87±.006 & 2.01±.003 & 7.23±.006 & 0.58±.002 & 1.85±.003 & 1.01±.002 & 3.06±.004 \\ \addlinespace[0.2em]
FAISS   & \underline{2.24±.003} & \underline{7.19±.006} & \underline{2.11±.003} & \underline{7.54±.006} & \underline{0.72±.002} & \underline{2.49±.004} & \underline{1.07±.002} & \textbf{3.75±.004} \\ \addlinespace[0.2em]
ComEM   & 1.98±.003 & 6.56±.005 & \textbf{2.22±.003} & \textbf{7.84±.006} & 0.60±.002 & 2.30±.003 & 1.03±.002 & 3.47±.004 \\ \addlinespace[0.2em]
GNEM    & \textbf{2.25±.003} & \textbf{7.30±.006} & \textbf{2.22±.003} & \textbf{7.84±.006} & \textbf{0.77±.002} & \textbf{2.54±.003} & \textbf{1.11±.002} & \underline{3.70±.004} \\
\bottomrule
\end{tabular}
\vspace{-0.8em}
\end{table*}

Our downstream evaluation in Table~\ref{tab:retrieval} reveals interactions between EM methods and LLM capabilities in conversational recommendation systems. 

From EM perspective, the performance hierarchy observed in Reddit-Amazon-EM evaluation that GNEM > ComEM > traditional methods persists but attenuates in conversational settings. GNEM retains its lead with the highest R@5 scores across all LLMs, achieving 7.30\% for GPT-4 and 7.84\% for GPT-3.5, thereby demonstrating superior robustness to conversational variations. The absolute performance gaps between the methods significantly narrow compared to the Reddit-Amazon-EM benchmark. This indicates that LLM outputs and conversational noise considerably impact the performance of CRS.

From the LLM perspective, GPT-4 achieves the highest R@5 of 7.30\% with GNEM, while smaller LLMs such as Qwen3-4b and Phi3-mini show significant degradation, with a 1.85-3.75\% R@5. This demonstrates that 1) Powerful LLMs produce more structured references that better align with EM methods' matching patterns; 2) EM methods enhance CRS effectiveness through precise grounding, as accurate entity matching converts LLM-generated references into database-aligned items, enabling contextually relevant recommendations that maintain dialogue coherence. The performance gap between GPT-4 and Qwen3-4b with GNEM in R@5 underscores that smaller LLMs generate noisy entity mentions that defy even robust matching methods.

These findings reconcile with our benchmark results through two lenses. First, though attenuates in conversational settings, EM methods' general performance hierarchy on Reddit-Amazon-EM that GNEM > ComEM > Hybrid > Traditional persists, demonstrating that our benchmark metrics reflect upper-bound performance of entity matching methods. Second, the precision-recall tradeoff observed in benchmark evaluation , namely Faiss' high recall but low precision, manifests differently here. Faiss achieves R@5 comparable  to GNEM on GPT-3.5 in R@5, suggesting that conversational breadth and variation in mentioning movie titles may compensate for matching inaccuracy. 

%% file: contents/6_conclusion.tex
\section{Conclusion}

In this paper, we present Reddit-Amazon-EM, the largest publicly available knowledge-grounded dataset for entity matching featuring over 4k manually verified movie mappings between Reddit conversations and Amazon's structured catalogue. Reddit-Amazon-EM enables rigorous evaluation of EM methods in recommender systems. Our experiment on Reddit-Amazon-EM shows that graph-based, LLM-enhanced and hybrid EM methods demonstrate significant performance improvements over traditional baselines. Furthermore, our experiments on the downstream task of LLM-driven CRS dialogues confirmed the utility of Reddit-Amazon-Movies dataset as a robust benchmark for future research. By releasing the dataset, annotated mappings, and evaluation code, we aim to foster reproducible advancements in entity matching to facilitate future research in LLM-based conversational recommendations and knowledge-grounded dataset construction. This work not only advances the field of entity matching but also paves the way for more dynamic and context-aware recommender systems.

\newpage

\section{Limitations}
Our work focuses on establishing a high-quality benchmark for rigorous EM through careful manual annotation. While this ensures high reliability of data, the annotation process is inherently resource-intensive. Exploring efficient strategies for scaling annotation or leveraging weak supervision for similar tasks warrants future investigation.